\documentclass[twocolumn,floatfix,aps,prb,eqsecnum,superscriptaddress]{revtex4}

\usepackage{graphicx}
\usepackage{amsmath}
\usepackage{amssymb}
\usepackage{bm}
\usepackage{color}
\usepackage{bbold}
\usepackage{mathtools}

\begin{document} 
\title{General scheme for stable single and multiatom nanomagnets according to symmetry selection rules}
\author{M. Marciani}
\affiliation{Instituut-Lorentz, Universiteit Leiden, P.O. Box 9506, 2300 RA Leiden, The Netherlands}
\author{C. H\"{u}bner}
\affiliation{I. Institut f\"{u}r Theoretische Physik, Universit\"{a}t Hamburg, Jungiusstr.  9, 20355 Hamburg, Germany}
\author{B. Baxevanis}
\affiliation{Instituut-Lorentz, Universiteit Leiden, P.O. Box 9506, 2300 RA Leiden, The Netherlands}

\begin{abstract}
At low temperature, information can be stored in the orientation of the localized magnetic moment of an adatom. However, scattering of electrons and phonons with the nanomagnet leads its state to have incoherent classical dynamics and might cause fast loss of the encoded information. Recently, it has been understood that such scattering obeys certain selection rules due to the symmetries of the system. 
By analyzing the point-group symmetry of the surface, the time-reversal symmetry and the magnitude of the adatom effective spin, we identify which nanomagnet configurations are to be avoided and which are promising to encode a stable bit.  A new tool of investigation is introduced and exploited: the \textit{quasi}-spin quantum number. By means of this tool, our results are easily generalized to a broad class of bipartite cluster configurations where adatoms are coupled through Heisenberg-like interactions.  Finally, to make contact with the experiments, numerical simulations have been performed to show how such stable configurations respond to typical scanning tunneling microscopy measurements.

\end{abstract}
\maketitle

\section{Introduction}

In recent years, great effort has been made to scale down the dimension of spintronic devices able to store classical bits of information. For this purpose, current research is devoted to understand the physics of single atoms and small clusters absorbed on non-magnetic metallic\cite{Gambardella2003,Khajetoorians2011,Khajetoorians2013,Miyamachi2013} or insulating\cite{Otte2008,Loth2012,Rau2014,Loth2010,Donati2016} surfaces.
The theoretical description of the dynamics of such systems is challenging as it lies at the intersection of classical\cite{Apalkov2005,Li2004,Taniguchi2013} and quantum\cite{Gatteschi2006} mechanics.

The low temperature dynamics of suitable adatoms, without applied magnetic field, may be described by two degenerate low-energy states with opposite magnetization. These states can be naturally regarded as the bit constituents. Unfortunately, not all adatoms present this feature as it relies on specific environmental conditions like the hybridization mechanism with the surface and the symmetry of the crystal field produced by the substrate\cite{Steinbrecher2016,Donati2014}. In particular, some systems exhibit no degenerate groundstate and the two lowest-energy states have no magnetization at all. This feature is referred to as groundstate splitting (GSS) and is due to the coupling of the orbital degree of freedom of the adatom with the crystal field.

To be suitable as memory storage\cite{Kalff2016}, an engineered bit is required to retain its state over an extended time period\cite{Khajetoorians2016}. Hyperfine interactions inside the adatom\cite{Wernsdorfer2000} and the contact with the substrate induce the atomic state to have an incoherent dynamics. In particular, the scattering of electrons and phonons off the adatom may be such that the stability of its state is affected drastically due to frequent switching between the groundstates.

With time the scientific community has started to recognize the role played by the symmetries of the system\cite{Miyamachi2013,Khajetoorians2016,Hubner2014}. Their implications are extremely relevant not only in determining whether the two low-energy atomic states are magnetized but also in constraining their stochastic dynamics. In particular, first order processes mediated by the substrate electrons that make the adatom in one low-energy state to jump to another one - usually called single-electron (SE) switching processes - may be inhibited by symmetry selection rules\cite{Delft1992}. However, symmetry information alone is not always sufficient. According to models currently in use\cite{Gatteschi2006,Bartolome2014}, it must be contrasted with the magnitude of the effective total angular momentum of the adatom.

\begin{figure}
\includegraphics[width=.35\textwidth]{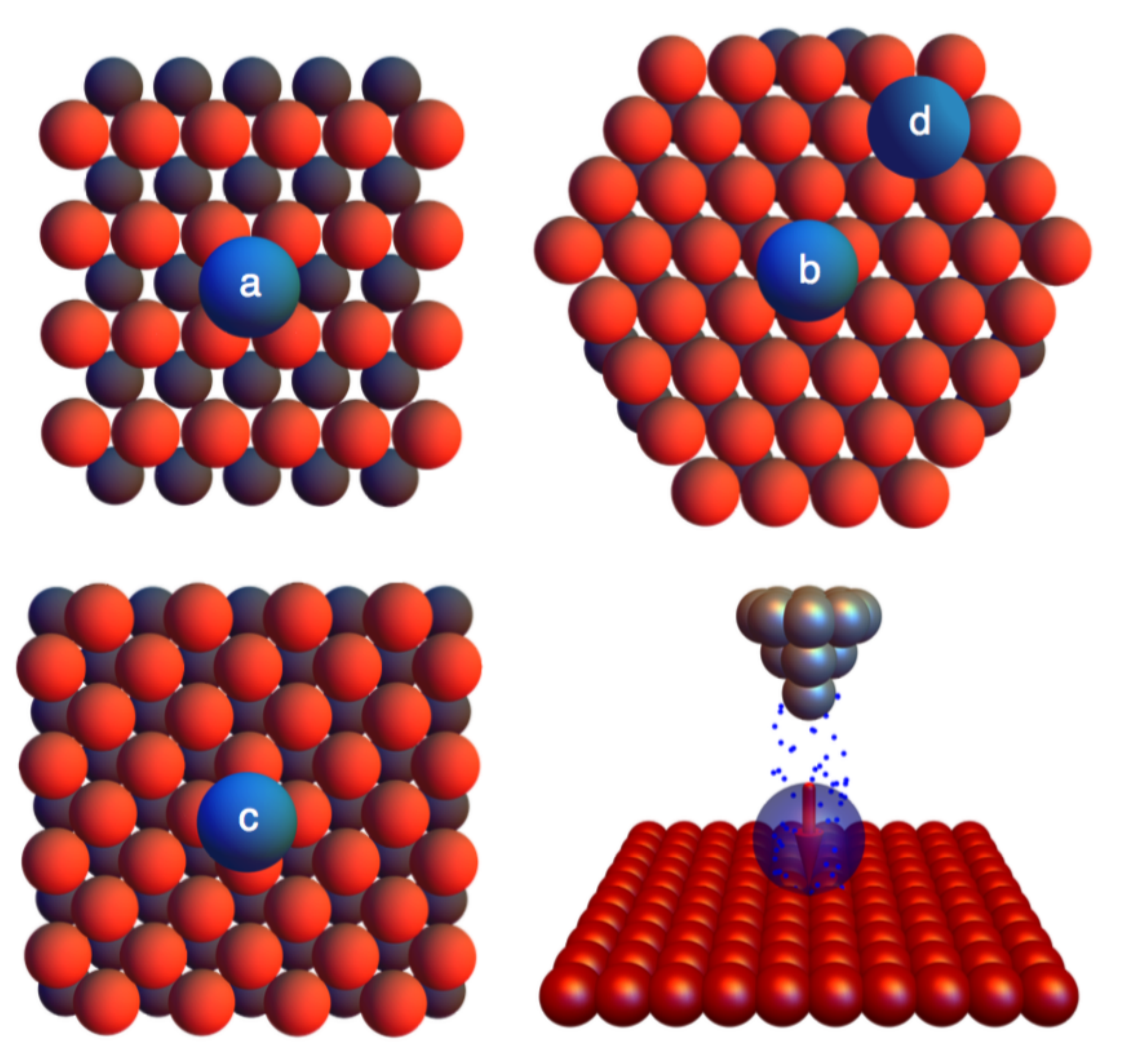}
\caption{\textit{(a)-(d)} Atoms deposited on different surfaces with $C_{\chi v}$ symmetry. $\chi = 2,3,4,6$, respectively for the adatoms (a),(b),(c),(d). (\textit{bottom right}) Sketch of a scanning tunneling microscope current measurement to infer the total momentum of the adatom. The tip of the microscope (in grey) exchanges electrons with the surface through the adatom.} \label{fig:image1}
\end{figure}

In this paper we present a general scheme to explain and predict exceptional long lifetimes of spin orientation in single and multi atomic systems. Hereby we provide a complete and rigorous map of such combinations of symmetries and total angular momentum magnitude, valid for small transversal crystal field. The symmetries we consider are the spatial point group $C_{\chi v}$ of the surface (see Fig.~\ref{fig:image1}) and time-reversal. We consider the possibility that the time-reversal symmetry could be broken by a finite magnetic field perpendicular to the surface. Our findings are in agreement with existing experimental\cite{Miyamachi2013,Donati2016} results and previous numerical\cite{Hubner2014} and analytical\cite{HuebThesis} studies.  With the restriction to time reversal symmetry a classification scheme\cite{Prada2017} was presented, which is related to a non-trivial geometric phase. However, we noticed a difference in the prediction of stable systems in the common case of zero magnetic field.

Further, we generalize our findings to multiatom clusters where adatoms are coupled with each other via bipartite Heisenberg interactions. This extension creates also a link between our work and classical research on general properties of spin systems\cite{Lieb1962,Nachter2007}.

\section{Single atom nanomagnet} \label{sec:satom}

\subsection{Model}
The Hamiltonian we consider can be decomposed as summation of parts related to the atom ($A$), to the electrons in the substrate ($S$) and their mutual interaction
\begin{align} \label{totalHam}
 H = H_{A} + H_S  + H_t.
\end{align}
The atom is assumed to be described, at low temperature, by a magnetic moment of magnitude $J$. For instance, this is the case of some rare-earth atoms\cite{Wybourne1965}, whose strong internal spin-orbit coupling is such that only one multiplet of the total angular momentum plays a role in the low energy physics, and transition metal ions\cite{Abragam1970}. The atom, affected by the substrate crystal field and subject to an external magnetic field $\vec B$, can be described by the single-spin Hamiltonian 
\begin{equation} \label{HA}
H_{A} =  H_{A}^{(0)} + H_{A}^{(1)} + \vec{B} \cdot \vec{J}, \quad H_{A}^{(0)} = - \vert D \vert \,J^2_{z} 
\end{equation}
where $H_{A}^{(0)}$ represents the so-called uniaxial (longitudinal) anisotropy (at second order) and $H_{A}^{(1)}$ contains higher order uniaxial and transversal anisotropy terms. The coefficient $|D|$ has been found as big as $1.5\,meV$ in Fe deposited on CuN\cite{Hirjibehedin2007} and $0.1\,meV$ in Fe deposited on Cu(111)\cite{Khajetoorians2013}.
In the rest of the paper we will refer to $J$ as a spin degree of freedom for brevity; however, the reader must intend that we mean total angular momentum. The substrate Hamiltonian is that of a single-band metallic Fermi liquid with no self-interactions:
\begin{align}
 H_S = \sum_{k,\sigma} \varepsilon_k \,\, c^{\dagger}_{k,\sigma} \,  c_{k,\sigma}.
\end{align}
Finally, we describe the effective interaction between metal and adatom by the Appelbaum Hamiltonian\cite{Appelbaum1967}
\begin{align} \label{Ht1}
 H_t = \kappa \, \vec J \cdot \vec j
\end{align}
where $\kappa$ is a momentum-independent coupling strength and $\vec j=  c^{\dagger}_{x=0} \vec \sigma c_{x=0} \propto \sum_{k,k'} c^{\dagger}_{k} \vec \sigma c_{k'}$ is the effective spin degree of freedom of the metal electrons coupled to the atom. Here and later $\sigma_i$ are the Pauli matrices and $\hbar=1$. 

We assume the temperature to be large enough, to justify a perturbative master equation approach\cite{Timm2008} and neglect strong correlations with the bath, such as the Kondo effect or energy renormalization\cite{Delgado2015}. On the other hand, thermal excitations should be small enough to ensure only the ground states to be occupied and resemble switching dynamics of a two level system. According to the Boltzmann distribution, the temperature should verify $k_b T \lesssim 0.1 \Delta$, where $\Delta \propto |D|$ is the energy gap between the two lowest-energy levels and the other ones. We will not treat atomic hyperfine interactions.

\subsection{Operators} \label{sec:opera}

Three physical operations on the system are relevant for our analysis of the stability of the atomic nanomagnet: rotation with discrete angles with axis perpendicular to the surface, time reversal (TR) and mirror across a certain mirror plane. We define here their representations in the atomic spin space. In the next sections, we will regard these operations as symmetries of the atomic system and analyze the consequences on the stability of the groundstate.\\

\paragraph{Rotation generator.}
The rotational symmetry of the adatom within the crystal field maps onto a rotational symmetry into the spin space. The generator of the rotation group is represented by
\begin{align} \label{rotdef}
R_{z,2\pi/\chi}  = \exp\{i\frac{2 \pi}{\chi}\,J_z\}.
\end{align}
The rotation generator has the property $R^\chi = \pm 1$ (we will omit the subscript in $R_{z,2\pi/\chi}$ for the rest of the paper), where the plus refers to integer spin systems and the minus to half-integer ones. This generator has at most $\chi$ distinct unit eigenvalues, equal to $r_{\chi}=\exp\{ i\,2\pi n/\chi \}$ with $n \in \mathbb{Z}$, for integer momentum systems, and $n \in \mathbb{Z}+1/2$, for a half-integer ones. \\

\paragraph{Time reversal operator.}

Time reversal is represented by the antiunitary operator

\begin{align} \label{Trev}
T  = \exp\{i\pi\,J_y\}\,K,
\end{align}

acting on the basis $\{|J,j_z \rangle \}$, where $K$ is the conjugation operator. In the following we will shorten the notation of the basis states as $\{|j_z \rangle \}$.

The action of $T$ can be defined such that $T |j_z \rangle = (-1)^{\lfloor{j_z\rfloor}} |-j_z \rangle$, where $\lfloor{\cdot \rfloor}$ is the floor function. The square of the TR operator acting on a integer or half-integer momentum Hilbert space gives $1$ or $-1$, respectively\cite{Wigner1959}. 

$T$ commutes with $R$. Nonetheless its antiunitarity hinders the possibility to find a common eigenbasis. Indeed, suppose $|\psi\rangle$ is an eigenstate of $R$ with eigenvalue $r$, then $TR\, |\psi\rangle=T\,r\,|\psi\rangle=r^*\,T\,|\psi\rangle$. At the same time $TR\,|\psi\rangle=R\,T|\psi\rangle$ and we conclude that $T|\psi\rangle$ is an eigenstate of $R$ but with eigenvalue $r^*$. Considering the quantity $\langle T \psi| R |\psi\rangle$ and applying $R$ in the bracket first to the left and then to the right state, one immediately concludes that $T|\psi\rangle \perp |\psi\rangle$ when $r$ is non real. Only if $r$ is real we can find a $|\psi\rangle$ which is eigenstate of both $T$ and $R$. We will use this feature later, in section \ref{sec:GSSsingle}.

In other words, even though two commuting symmetries are present, eigenstates cannot be in general labeled with two well defined quantum numbers at the same time.\\
 
\paragraph{Mirror operator.}

Freedom in choosing the coordinate axes allows to set one mirror plane along $yz$. We call $M$ the operator that reflects across this plane. Then, all other possible reflections with the other mirror planes are constructed conjugating it with the elements of the rotation group.

Since $\vec J$ is a pseudo-vector, $M$ acts on the spin fundamental algebra transforming $J_{y,z}$ to $(-J_{y,z})$ while keeping $J_{x}$ unchanged.  To obtain the explicit representation, we notice that this operator is equivalent to a $\pi$ rotation around $x$. Therefore,
\begin{align} \label{Rm}
M = e^{i\pi\,J_x}.
\end{align} 
Notice that $M^2=\pm 1$ (the plus refer to integer spins systems and the minus for half-integer ones) and that $M\,R = R^\dagger\,M$.

\subsection{Hamiltonian symmetry constraints and Stevens operator expansion} \label{sec:stev}

Using all symmetries we can characterize the most general structure that the Hamiltonian can have. In Ref. \citenum{Wybourne1965} a general tesseral harmonic expansion of $H$ compatible with a number of point symmetry groups is discussed and relative constraints are found. Here, we stick to the point group $C_{\chi\,v}$ symmetry and analyze the Stevens operator expansion of the Hamiltonian $H_A$ in Eq. \eqref{totalHam}. We start considering the spatial symmetries constraints, then we show the one due to the TR symmetry.

A generic Stevens operator\cite{Stev1952} $O^q_p$ (with $q<p$) is expressed in a closed form in Ref.  \citenum{Ryabov1999}. These operators are Hermitian by construction and, after trivial manipulations, we can write them in the following form:
\begin{align} \label{StevOp}
O^q_p  & =  \frac{1}{2}\sum_{r=0}^{\left \lfloor{(p-q)/2}\right \rfloor} c(p,q,r) \left\{ J_+^q +  J_-^q, J_z^{p-q-2r}\right\}, \nonumber \\
O^{-q}_p & = \frac{i}{2}\sum_{r=0}^{\left \lfloor{(p-q)/2}\right \rfloor} c(p,q,r) \left\{ J_+^q - J_-^q, J_z^{p-q-2r}\right\},
\end{align}
where $q$ and $p$ are natural numbers and $c(p,q,r) $ are real prefactors whose magnitude is not relevant for our discussion.

Since the atomic system has spatial symmetry $C_{\chi v}$, the equations
\begin{align} \label{Rsymmetry}
[H_A, R] &=0, \nonumber\\
[H_A, M] &=0
\end{align}
must hold.

The first equation implies that all matrix elements of $H$ between states with different eigenvalue $r_\chi$ must vanish. Moreover, we can expand $H_A$ using the operators in Eq. \eqref{StevOp}. Each operator $O^q_p$ or $O^{-q}_p$, when applied to the basis state $|j_z\rangle$, transforms it to a superposition $\alpha|j_z+q\rangle + \beta|j_z-q\rangle$. The superposition retains the rotation eigenvalue of the latter state only if $r_\chi(J_z\pm q)=r_\chi(J_z)$ i.e. if $q= m\chi, m\in \mathbb{N}$\cite{note1}. Therefore, only terms proportional to $O^{\pm m \chi}_p$, are allowed in the expansion. 

Notice that rotational symmetry in our problem is analogous to translation symmetry in one dimensional periodic crystals. The Hamiltonian eigenstates can be labeled with their eigenvalues $r$ and the latter are in one to one correspondence with a set of \textit{quasi}-spin\cite{note2} defined in a one dimensional Brillouin zone (BZ). Such a set is isomorphic to $\mathbb{Z}_\chi$ and can be defined as $\{-\lfloor \chi/2\rfloor+1,-\lfloor \chi/2\rfloor+2,\dots,\lfloor \chi/2\rfloor \}$, for systems with integer $J$, and $\{-\lceil \chi/2\rceil+1/2,-\lceil \chi/2\rceil+3/2,\dots,\lceil \chi/2\rceil-1/2\}$ for systems with half-integer $J$ (notice the use of floor and ceiling functions here). For instance, for half-integer spin systems with $\chi=3$ the BZ is $\{-1/2,1/2,3/2\}$; for integer ones with $\chi=4$, the BZ is $\{-1,0,1,2\}$.  Clearly, every spin state has a well defined \textit{quasi}-spin in the above defined BZs and this is equal to 
\begin{align} \label{qmom}
J^{(q)}_{J_z} \vcentcolon =  \left( \left[J_z+(\chi-1)/2 \right] \bmod \chi \right) \,-\,  (\chi-1)/2.
\end{align}
where we make use of the modulo operation ($x \bmod y$ indicates the value of $x$ modulo $y$).

\begin{figure}
\includegraphics[width=.5\textwidth]{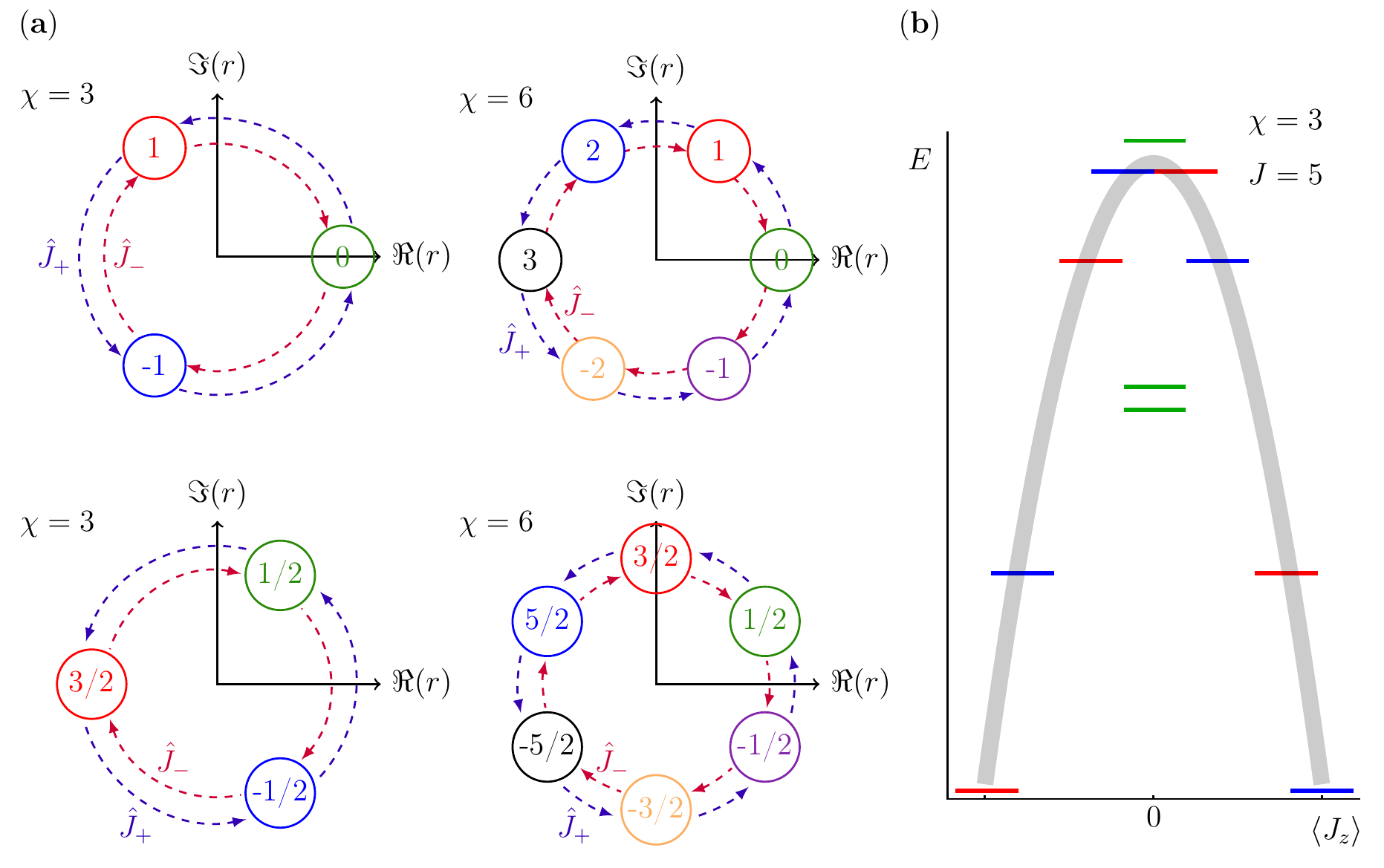}
\caption{(\textit{a}) Periodic Brillouin Zones (BZs) for integer spin systems (top) and half-integer ones (bottom). To better visualize the periodicity of the BZs, their elements (the little circles) are placed at the complex eigenvalues of $R$ and the number they contain indicates the associated \textit{quasi}-spin. Blue(red) arrows indicate SE transitions with transfer of positive(negative) \textit{quasi}-spin. (\textit{b}) Typical spectrum of a three-fold rotation symmetric system with small transversal anisotropy. On horizontal axis is the average magnetization along $z$ of the levels. The color code of the level indicates its \textit{quasi}-spin according to the top left case in (a). All figures are adapted from Ref. \citenum{HuebThesis}.} \label{fig:circles}
\end{figure}

For instance, the spin state with $J_z=-4$ in a system with $\chi=3$ has $J^{(q)}=-1$.
More ``bands" are present as soon as $J\geq \chi/2$ i.e. when $J$ is such that at least two different spin states have the same \textit{quasi}-spin. Fig.~\ref{fig:circles}(a) shows the periodic BZs for $\chi=3,6$.

The mirror operator $M$ acts with the transformations $\left(J_z,J_{\pm}\right) \rightarrow \left(-J_z,J_{\mp}\right)$. Eq. \eqref{Rsymmetry} implies $[M,O^{\pm q}_p]=0$ and the latter equation constrains the difference $p-q$ to be even(odd) when the superscript of $O$ is positive(negative). Hence, combining this constraint with the rotational one, we see that only operators of the form $O^{{m\chi}}_{{m\chi}+2n}$ and $O^{-{m\chi}}_{{m\chi}+2n+1}$ with $m,n\in \mathbb{N}$ are allowed.

Finally, TR operator acts with the transformation $(J_{\pm},J_z) \rightarrow - (J_{\mp},J_z)$ and $i \rightarrow (-i)$. Consequently, TR  symmetry, if present, implies the label $p$ to be even.

To be explicit, when all symmetries are present, the allowed Stevens operators in the expansion of $H_A$ only $O^{(-1)^{m\chi} m \chi}_{2n},\;(m,n \in \mathbb N)$. Notice that the Hamiltonian would be always real (in the spin eigenbasis $\{|j_z \rangle\}$) for $\chi\neq 3$, but is in general not real for $\chi=3$\cite{note3}. \\

In the following, we will use the \textit{quasi}-spins as quantum numbers to label the atomic eigenstates. In some cases, the eigenvalues of the mirror operator $M$ could be added to the set of the quantum numbers. However, its eigenstates present no magnetization along the $z$ direction\cite{note4} and are not suitable for the analysis of the next sections. Thus, the rotational symmetry is a central ingredient in determining the stability of the nanomagnet.

In the rest of the paper we will allow also for TR symmetry breaking due to magnetic field. However, only the component $B_z$ is allowed as is the only one which preserves rotational symmetry. \textit{Per contra}, the mirror symmetry gets broken. Notice that the antiunitary product operator $TM$ would still represent a symmetry for the system. We have checked the implications of this symmetry.  It is antiunitary and surprisingly allows for an additional quantum number for the Hamiltonian eigenstates. However, since it does not provide strong selection rules for GSS or SE switching processes, we limit ourselves to briefly mention them in App. \ref{Rmconstraint}.

\subsection{Groundstate Splitting at $H_t=0$} \label{sec:GSSsingle}

We now turn our attention to the first goal: to show that, \textit{assuming $H_t=0$ and $\vec B=0$, it is possible to tell whether the groundstate of the atom is degenerate or it is allowed not to be, only by knowledge of the symmetries and the magnitude $J$ of its spin}. 
  
First, switch off momentaneously $H_{A}^{(1)}$ in $H$ (with $H_t=0$ and $\vec B=0$). The two degenerate groundstates are $|\psi_{GS}\rangle \vcentcolon =  |j_z=J\rangle$ and $|\tilde\psi_{ {GS}}\rangle \vcentcolon =  T |\psi_{GS}\rangle \propto |-J\rangle $ (we will omit '$j_z=$' for the rest of the paper). Even though $H_{A}^{(0)}$ has symmetry $C_{\infty v}$, it is convenient to identify already their eigenvalues under the action of the rotation generator $R_{z,2\pi/\chi}$ (where $\chi$ is defined as the maximum value for which $[H_{A}^{(1)},R_{z,2\pi/\chi}]=0$ holds). They are $r_{GS} = (r_{\widetilde {GS}})^*= \exp \{ i \,J\, 2\pi/\chi \}$ ($r_{\widetilde {GS}}$ is the eigenvalue for $|\tilde\psi_{ {GS}}\rangle$) and their \textit{quasi}-spin are defined in Eq. \eqref{qmom}. \\
Now, we switch on $H_{A}^{(1)}$ adiabatically to its actual value. Energies and eigenstates change along the process, but the \textit{quasi}-spin of all eigenstates are preserved since $[H_{A}^{(1)},R]=0$. At the end of the process the groundstates of the system would have retained their initial \textit{quasi}-spins unless some state with different \textit{quasi}-spin crossed the groundstates along the process, becoming lower in energy. Since $H_{A}^{(1)}$ is left generic in our analysis, we can not have control on the final value of the groundstate \textit{quasi}-spin after such crossings. To prevent these inconvenience, we assume $H_{A}^{(1)}$ to be small enough (roughly speaking, $ H_{A}^{(1)} \ll H_{A}^{(0)}$ is sufficient).\\

Using the properties of the TR operator illustrated Sec. \ref{sec:opera}, we claim that \textit{eigenstates $|\psi\rangle$ of both $H_A$ and $R$ with non-real $r$ are degenerate in presence of TR symmetry}. \\

Clearly, this statement is non-trivial only for integer spin systems because half-integer spin ones under TR symmetry always exhibit groundstate degeneracy by Kramers theorem.
To prove the claim, remind that if $r$ is non-real then $|\tilde \psi \rangle \vcentcolon =  T|\psi\rangle \perp |\psi\rangle$. Subsequently, $[H,T]=0$ implies that, on one hand $TH|\psi\rangle=\varepsilon_0 T |\psi\rangle = \varepsilon_0 |\tilde \psi\rangle$ and on the other hand $TH|\psi\rangle= H T|\psi\rangle= H |\tilde \psi\rangle$. Hence, joining together the two equations, we get $H\,|\tilde \psi\rangle= \varepsilon_0 |\tilde \psi\rangle$.\\

The statement above applies to the groundstate. We conclude that it can get split by tranversal anisotropy terms only if $r_{GS}$ is real or, in other words, if its associated \textit{quasi}-spin is a TR invariant point of the Brillouin zone ($|J^{(q)}_{GS}| = - |J^{(q)}_{GS}|+m\chi, m  \in \mathbb{N}$). Thus, the splitting happens when
\begin{align} \label{GSS2contraint}
\exists m  \in \mathbb{N} \,:\, J = \frac{m\chi}{2}.
\end{align}  
This constraint determines the columns GSS in the Tabs. \ref{table_bos} and \ref{table_fer}.
When the system features GSS in presence of TR symmetry, the two lower states are also non magnetic. They have to be eigenstates of the TR operator, therefore, $\{J_z,T\}=0$ implies $\langle \psi_{GS} | J_{z} | \psi_{{GS}} \rangle = 0$. We stress that the splitting may be also seen as a consequence of lowering the symmetry from the $C_{\infty v}$ subgroup of the free atom point group to the $C_{\chi v}$ subgroup of the atom within the crystal field.

\subsection{Single-electron switching process at $H_t \neq 0$} \label{SESsingle}

Finally, we switch on the interaction with the metal, $H_{t} \neq 0$. When the substrate gets coupled with the atom, the energy and \textit{quasi}-spin of the atomic state are not preserved anymore, because of scattering with the metal electrons. Since the metal has many degrees of freedom with respect to the atom, it is usually assumed to thermalize quickly and its Boltzmann distribution, being a classical one, leads the atom to have also an associated classical distribution\cite{Cohen1998}. The approximated Markovian law, that describes the dynamics of energy-defined states of the atom (the pointer basis of the nanomagnet \cite{Zurek2003}), is well known in literature \cite{Fernandez2009,Delgado2010}. However, there is an ambiguity in the definition of the pointer basis when the atom presents pairs of degenerate states (which is the case when the atom has no GSS and applied magnetic field). There are indications\cite{Delgado2016b} that the states of the pointer basis are those with maximum magnitude of the average magnetization, as the dephasing due to the scattering is the largest for these states. Thus, we are allowed to assume that the pointer basis coincides with the atomic eigenstates considered in the previous sections, with well defined \text{quasi}-spin.

It was shown\cite{Delgado2015} that the GSS feature might be destroyed when the Kondo coupling times the substrate electronic density of states gets large via a mechanism of gap 
quenching. However, such a mechanism is not effective in most of the experiments performed, therefore here we limit the discussion to small Kondo couplings i.e. $H_{t} \ll H_{A}$.

The rate of switching between two atomic eigenstates, say $|\psi_a \rangle$ and $|\psi_b\rangle$, at lowest order in $H_t$, i.e. due to a SE scattering with the atom, is
\begin{align} \label{gamma}
\Gamma_{ab} &= \frac{2\pi\kappa^2}{\hbar}\sum_{\mu,\nu} | \langle  \psi_a,\nu | H_t | \psi_b,\mu\rangle|^2 \, e^{-\beta E_\mu} \delta (x) \nonumber \\
&= \frac{2\pi\kappa^2}{\hbar} \sum_{\mu,\nu}\, \left\vert \sum_{s\in \{+,-,z\} }\!\!\!  \langle  \psi_a| J_{s} | \psi_b \rangle\, \langle  \nu| j_{\bar s}| \mu \rangle \right\vert^2 \, \!\!\! e^{-\beta E_\mu} \delta (x) 
\end{align}
where $\mu,\nu$ are states in the substrate, the bar in $j_{\bar s}$ indicates that the subscript takes opposite sign if $s=\pm$ and $x=E_{\nu}-E_{\mu}+E_{a}-E_{b}$.
It is clear that transitions are possible only when the states are connected by an operators $J_{s}$, with $s={+,-,z}$.

We show that the rotational symmetry provides  a selection rule on SE switching processes. The commutation relations between $J_s$ and $R$ are $RJ_{s}=e^{ i\varphi_s} J_{s}R$, where $\varphi_s = 0,\pm 2\pi/\chi$ respectively for $s=z,\pm$. Since the states $\psi_{a,b}$ are also eigenvalues of $R$, one gets:
\begin{align} \label{selrule}
 \left[ e^{i(\varphi_b - \varphi_a + \varphi_s)} -1 \right] \langle \psi_a | J_s | \psi_b \rangle = 0.
\end{align}
Thus, given $\psi_{a,b}$, at most one value of $s$ is such that $\varphi_s = \varphi_a - \varphi_b$. This means that a SE transition produces a \textit{quasi}-spin change equal to either $0,1$ or $-1$. When the \textit{quasi}-spins of the states differ by more than one, we are guaranteed that $\Gamma_{ab}=0$ and there is no SE transition between the two states. For instance, systems with $\chi=6$ and $J=15/2$ have groundstates with $J^{(q)}=\pm 3/2$ therefore at least three SE transitions are needed for a groundstate switching. One could easily check it using Fig.~\ref{fig:circles}(a) (SE transitions from the eigenstates are shown with arrows).

A second selection rule comes from the TR symmetry. It protects degenerate groundstates of integer spin systems from SE switching. Given $|\psi_{ {GS}} \rangle$ and $| \tilde \psi_{ {GS}} \rangle$ as the two time-reversal groundstate partners and making use of $\{J_z,T\}=0$ and $J_+\,T=-T\,J_-$ one finds\cite{Miyamachi2013,Delft1992} for all $s\in \{+,-,z\}$
\begin{align} \label{TRSconstraint2}
\langle \psi_{GS} | J_{s} | \tilde \psi_{ {GS}} \rangle = 0\quad & \mathrm{\,for\;integer\; spin.}
\end{align}
Actually, this constraint is non-trivial only with $\chi=3$. In the other cases the groundstates are either already split by transversal anisotropy or have \textit{quasi}-spin difference greater than one. For instance, in the experimental set of Ref. \citenum{Hirjibehedin2007} (Fe atoms on CuN substrate with $J=2,\chi=2$) GSS is present and SE transitions between the two lowest-energy states are indeed observed even at $B=0$. 

Other weak constraints come from the mirror symmetry but they are not enough to make SE switching to vanish. We leave this discussion to App. \ref{Rmconstraint}.

As a final remark, we notice that also small spin systems with $\chi>2J>1$ are protected against SE switching process. This happens because there are no pairs of states with the same phase or, in other words, there is only one ``band" in the Brillouin zone. Only if $J=1/2$, the system groundstates can be connected by SE transitions.\\

\subsection{Suppression of SE switching process at $H_t \lesssim H^{(1)}_A \ll H^{(0)}_A$} \label{sec:supp}

As an application of the tools of analysis developed in the previous sections, we describe here a feature related to the suppression of SE switching rate in some systems, when the terms in $H^{(1)}_A$ gets \textit{uniformly} small.
We assume, therefore, that $ H_t  \lesssim H^{(1)}_A \ll H^{(0)}_A$, making the further assumption that the different prefactors in front of each  $J^n_s\; (n \geq 0;s=+,-,z)$, in the expansion of $H^{(1)}_A$, have all the same order of magnitude $\varepsilon \ll 1$.
In this regime we can treat $H^{(1)}_A$ as perturbation of the system with Hamiltonian $H^{(0)}_A$.

Consider now $\Gamma_{\psi_{GS},\tilde\psi_{GS}}$ in Eq. \eqref{gamma}, the transition rate of the SE switching process between the true groundstates. The groundstates can be expressed as a perturbation series in $\varepsilon$:
\begin{align}
| \psi_{GS} \rangle &= | J \rangle + \varepsilon \sum_{m} \alpha_m |J-m\chi \rangle  + \mathcal{O}\left( \varepsilon^2 \right) 
 \nonumber \\
|\tilde \psi_{GS} \rangle &\propto  |- J \rangle + \varepsilon \sum_{n} \alpha'_n |-J+n\chi \rangle  + \mathcal{O}\left( \varepsilon^2 \right) 
\end{align}
where $m(n)$ is a natural number such that $J-m(n)\chi>-J$ and $\{\alpha_{m(n)}\}$ are expansion coefficients\cite{note5}.

The quantity $\langle \psi_{GS} | J_{s} |\tilde  \psi_{GS}\rangle$ in $\Gamma_{\psi_{GS},\tilde\psi_{GS}}$ gets contributions of different perturbative orders, of the form $\varepsilon\, \alpha'_n \langle J-m\chi |J_s|-J \rangle$ or $\varepsilon \,\alpha_m \langle J|J_s|-J+n\chi \rangle$ and $\varepsilon^2 \,\alpha_m  \alpha'_n \langle J-m\chi |J_s|-J+n\chi \rangle$. We notice that, inside the sets of systems which exhibit SE switching, we can distinguish two subsets. The systems in the first one presents the $\mathcal{O}(\varepsilon)$ contributions while the systems in the second one not. The first subset contains systems in which the unperturbed groundstate $|-J\rangle$, call it the left one, has the same \textit{quasi}-spin of either $|J\rangle$ (in the half-integer case only) or $|J-1\rangle$. On the contrary, systems of the second subset possess a left groundstate which would have the same \textit{quasi}-spin of the state $|J+1\rangle$. Of course this state is not allowed, thus, the $\mathcal{O}(\varepsilon)$ contributions are vanishing. 
A systems falls in the second group when the difference between the \textit{quasi}-spin of $|\tilde  \psi_{GS}\rangle$ and $| \psi_{GS}\rangle$ (modulo $\chi$) is equal to one. The magnitude of its spin, then, must verify (we make use of Eq. \eqref{qmom})
\begin{align} \label{eqsupp}
\left(2J\right) \bmod \chi = \chi-1.
\end{align}
In this perturbative regime the SE switching rates are
\begin{align}
\Gamma_{\psi_{GS},\tilde\psi_{GS}} \propto
\begin{cases}
\kappa^2(\varepsilon^2 + \mathcal{O}(\varepsilon^3)\;) & \text{for the first subset,}\\
\kappa^2(\varepsilon^4 + \mathcal{O}(\varepsilon^5)\;) & \text{for the second subset.}
\end{cases}
\end{align}
where $\kappa \lesssim \varepsilon$ (the assumption $H^{(1)}_A \gtrsim H_t$ is to guarantee that the dominant switching path for the second subset remains the SE one and not a multiple-electrons one).
From this expression is clear how systems in the second subset have smaller SE switching rates in the perturbative limit. 
They are listed in the column ``Supp" in Tabs. \ref{table_bos} and \ref{table_fer}. 

\subsection{Numerical Simulations} \label{num1}

We demonstrate the consequences of the symmetry considerations on the switching rate of a single-atom nanomagnet when experimentally measured by spin-resolved scanning tunneling microscopy (STM). In previous experiments, the stability of few-atoms clusters was investigated by means of this technique\cite{Loth2012,Khajetoorians2013,Khajetoorians2016}. In particular, the switching rate between groundstates has been observed in the telegraph noise. Such an experimental setup can be described by adding the STM tip Hamiltonian to Eq. \eqref{totalHam} while accessible quantities like the bias voltage, temperature and external magnetic field are varied. For this purpose we solve the master equation (see Refs. \citenum{Khajetoorians2013,Hubner2014}) for a six-fold rotational symmetric system with small transversal anisotropy, $ H_{A}^{(1)} =\alpha_6^6 O_6^6$, and several different spin magnitudes. As already mentioned before, we neglect the small energy renormalization of the atomic levels due to the coupling with the tip. All rates will be given in units of the direct tunneling rate $\Gamma_0 = \pi v_S^4 (\rho_{T\uparrow} \rho_{S\uparrow} + \rho_{T\downarrow} \rho_{S\downarrow})$.

\begin{figure}[tb]
\centerline{\includegraphics[width=1\linewidth]{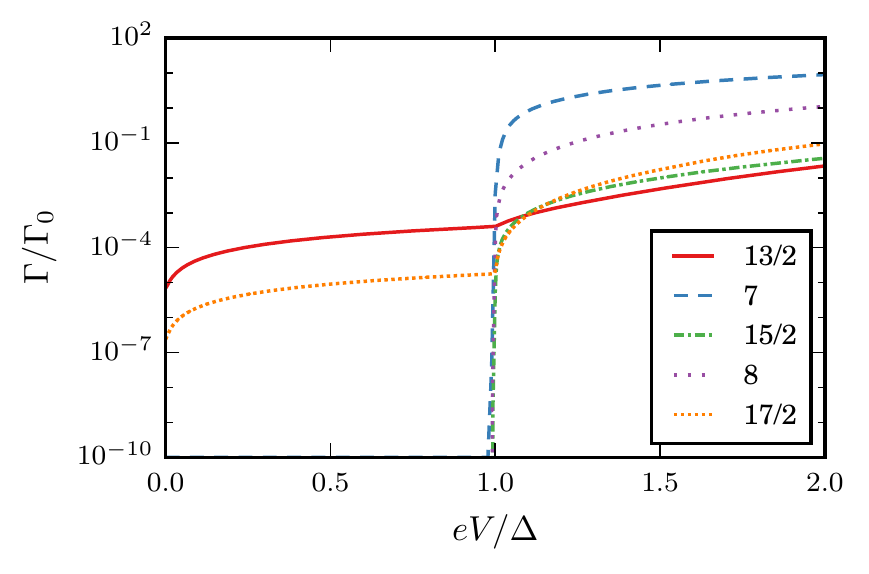}}
\caption{Bias-dependent switching rate of a spin with $J=13/2\ldots 17/2$ in a six-fold rotational symmetric crystal ($\chi=6$). Other parameters are $\kappa^2/D=0.1,\,\alpha_6^6/D=5\cdot 10^{-5}$, $kT/D=0.01$, $P=0.1$ ($P$ is the tip polarization) and $\Delta$ is the first excitation energy of the spin.}
\label{fig:bias}
\end{figure}

 Fig. \ref{fig:bias} shows the bias-dependent switching rate for several spin magnitudes. We observe that in all cases an increasing switching rate is observed for voltage higher than the spin excitation energy $\Delta$ of the magnet ($\Delta$ is the energy difference between the first excited state and the groundstate of the system with $B=0$). For the protected cases $J={7,\,15/2,\,8}$, however, the switching rate becomes negligible for low temperatures $kT\ll\Delta$ in accordance to Tabs. \ref{table_bos} and \ref{table_fer}. In contrast, $J=13/2$ and $17/2$ show SE switching even at low bias voltages resulting in a finite switching time $\tau = \Gamma^{-1}$. 

\begin{figure}[tb]
\centerline{\includegraphics[width=1\linewidth]{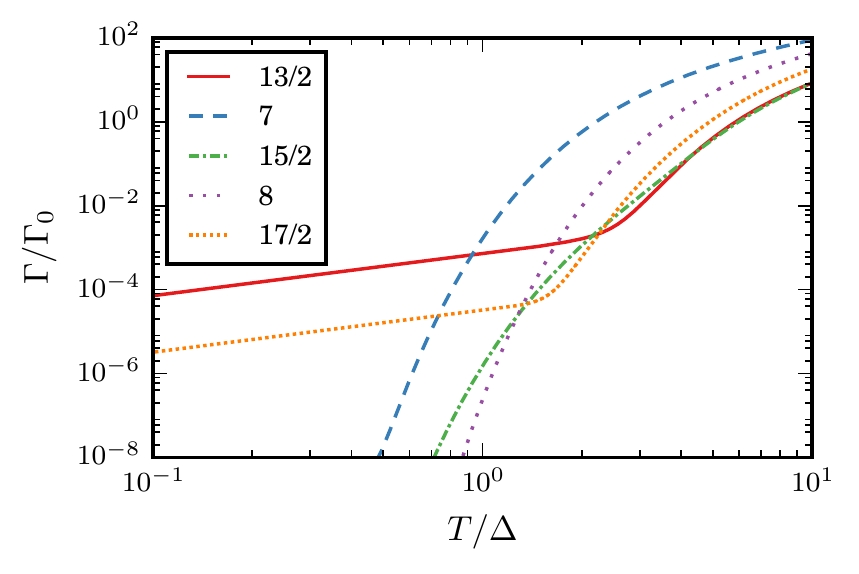}}
\caption{Zero-bias temperature dependency of the switching rate of a spin with $J=13/2\ldots 17/2$ in a six-fold rotational symmetric crystal ($\chi=6$). Other parameter as in in Fig. \ref{fig:bias}.}
\label{fig:temperature}
\end{figure}

Temperature-dependent switchings are investigated often by X-ray absorption spectroscopy and magnetic circular dichroism (XCMD) measurements to infer the stability of an atom or cluster (Fig. \ref{fig:temperature}). Similar to the bias-dependent measurement, one can observe, in all cases, an onset of the switching rate for temperatures high enough to excite the spin. At low temperature, the switching rate becomes negligible for the stable cases while remaining finite for unstable ones. In contrast to the bias dependency where the switching sets in abruptly at $eV=\Delta$ for stable atom configurations, the onset of the switching with temperature appears continuous and monotonously. 

\begin{figure}[tb]
\centerline{\includegraphics[width=1\linewidth]{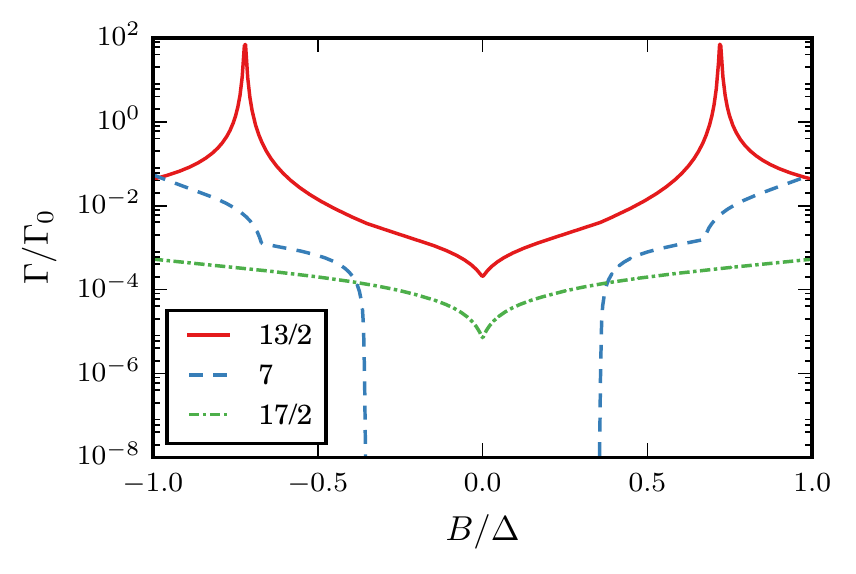}}
\caption{Magnetic field dependency of the switching rate of a spin with $J={13/2,\,7,\,17/2}$ in a six-fold symmetric crystal ($\chi=6$) for $eV/D=6$ and other parameters as in Fig. \ref{fig:bias}.}
\label{fig:field}
\end{figure}

In a next step, we break TR symmetry by applying magnetic field of strength $B$ along the $z$ axis (Fig. \ref{fig:field}). For the chosen magnetic field range, the cases $J=13/2$ and $J=17/2$ show SE switching as they are not protected by symmetry. In particular, $J=13/2$ shows a Lorentzian-like peak at the magnetic field strength at which one of the former groundstates gets degenerate with one of the former first excited states. The specific shape has to be associated to the fact that the two states have the same \textit{quasi}-spin and hybridize. In contrast, $J=7$ is stable for low magnetic field. However, spin switching gets activated at higher applied fields when the former groundstate is brought in resonance with one excited state. In this case the curve profile is different since the two states have different \textit{quasi}-spins.

\subsection{Discussion}

The results of our single-atom analysis are summarized in Tabs. \ref{table_bos} and \ref{table_fer}.

\begin{table}
\centering
\begin{tabular}{ | l|| c|| c | c|| c|| c| }
\hline
\qquad $\chi$ &  GSS & SES(T)  & SES(BT) & Protected & Supp\\ \hline \hline
\qquad $2$ & $\{ n\}$ & $\{ \}$ & $\{ n\}$ & $\{ \}$ & $\{\}$\\ \hline
\qquad $3$ & $\{ 3n\}$ &  $\{ \}$  &  $\{ n\}$ \textbackslash $\{ 1 \}$  & $\{1\}$ & $\{1+3n\}$\\ \hline
\qquad $4$ & $\{ 2n\}$ &  $\{ \}$ &  $\{ 2n\}$ & $\{1,3,5\}$ & $\{\}$\\ \hline
\qquad $6$ & $\{ 3n\}$ &  $\{ \}$ &  $\{ 3n\}$ & $\{1,2,4,5\}$ & $\{\}$ \\ \hline
\end{tabular}
\caption{Sets of integer spin magnitudes $\{J_n\}$, with $n\in \mathbb N_{>0}$, which exhibit groundstate splitting (GSS) or SE switching processes (SES), at given system symmetry $C_{\chi \, v}$. The etiquette ``(T)" and ``(BT)" differentiate on whether time reversal symmetry is, respectively, present or broken. ``$\{\}$" indicates the empty set and the notation ``$\{a\} \backslash \{b\}$" stands for the set subtraction of $\{b\}$ from $\{a\}$. The fourth column (Protected) shows instances of magnitudes which are protected from both GSS and SES.  The last column (Supp) shows the sets with suppressed SE switching processes at very small  $H^{(1)}$ and $H_t$, as described in Sec. \ref{sec:supp}.} 
\label{table_bos}
\end{table}

\begin{table}
\centering
\begin{tabular}{ | l || c | c || c || c | }
\hline
\qquad $\chi$ & GSS &  SES(T,BT)  & Protected & Supp\\ \hline \hline
\qquad $2$ & $\{ \}$  &  $\{ n\!+\!1/2\}$  & $\{\}$ & $\{ \}$  \\ \hline
\qquad $3$ & $\{ \}$ &  $\{ n\!+\!1/2\}$ \textbackslash $\{\frac{3}{2}\}$&  $\{\frac{3}{2}\}$ & $\{\frac{5}{2}\!+\!3n\}$ \\ \hline
\qquad $4$ & $\{\}$ & $\{ n\!+\!1/2\}$ \textbackslash $\{\frac{3}{2}\}$&  $\{\frac{3}{2}\}$ & $\{\frac{3}{2}\!+\!2n\}$ \\ \hline
\qquad $6$ & $\{ \}$ & $\{ n\!+\!1/2\}$ \textbackslash $\{\frac{5}{2},\frac{3}{2}\!+\!3n\}$  & $\{\frac{3}{2}, \frac{5}{2},\frac{9}{2},\frac{15}{2}\}$ & $\{\frac{5}{2}\!+\!3n\}$\\ \hline
\end{tabular}
\caption{Same as in Tab. \ref{table_bos}, but for half-integer spin magnitudes. Notice that TR symmetry does not provide additional protection from SE switching processes as it does in integer spin systems.}
\label{table_fer}
\end{table}

From our considerations, we can conclude that the higher the symmetry the more stable will be the bit encoded in the groundstates. To substantiate this statement we bring to the attention of the reader the cases of $\chi=2$ and $\chi=6$. The former case does not host good nanomagnets as either their groundstates are split or present SE switching processes. On the contrary, the latter case hosts nanomagnets with high stability against both SE and single-phonons switching processes\cite{note6}. Indeed, in half-integer spin systems with $J=\frac{3}{2}+3n,\;(n\in \mathbb{N})$ the  difference between the groundstates \textit{quasi}-spins is maximal, equal to 3.

We remark the advantage in working with the \textit{quasi}-spin formalism, analog to the \textit{quasi}-momentum formalism in crystal theory, in order to get universal formula for the presence of GSS and other features. The \textit{quasi}-spin would also be a more natural horizontal axis in typical spectrum plots encountered in literature, like the one in Fig.~\ref{fig:circles}(b). 

Notice that the mirror symmetry plays only a marginal role in our qualitative discussion: it does not provide strong constraints to GSS or SE switching processes. However, its inclusion is relevant for quantitative numerics where the correct (symmetry preserving) Stevens operators must be taken into account.

We warn the reader that our results refer to ``generic" Hamiltonians, that is, within a non-zero measure subset of the set of all possible symmetry preserving Hamiltonian. For example, a system with $J=9/2$ and $\chi=3$ would not present SE switching processes (in contrast with Tab. \ref{table_fer}) if only the Stevens operator $O_4^3$ is included in $H_A^{(1)}$. However, inclusion of higher order Stevens operators like $O_6^6$ would restore the agreement with our theory. 

The absence of SE switching processes in the case $J=3/2$ and $\chi=3$ is explained at the end of App. \ref{HamRep}. \\

As a final remark, we comment a few relevant, recent experiments.

One experiment is Ho on Pt(111) where the substrate has $3$-fold degeneracy. One experimental group\cite{Miyamachi2013} found the adatom spin magnitude to be $J = 8$ and measured low groundstates switching rate. According to our theory, such system would be protected from both GSS and SE switching if the transversal anisotropy is not too big (see Tab. \ref{table_bos}). The latter was actually computed by the authors by means of \textit{ab-initio} calculations. The ratio between the uniaxial anisotropy term and the biggest transversal anisotropy term was found to be approximately $0.1\%$. Such value is compatible with the absence of level crossing and allows the usage of our theory.
However, another experimental group\cite{Donati2014} found a strong fourth-order uniaxial term inducing a groundstate level crossing. The system groundstate then does not occupy the spin state $|J_z|=8$ anymore but rather it occupies the spin state $|J_z|=6$. In this case we can still use our theory in this way: the groundstate quasi-spin can be inferred using Eq. \eqref{qmom} with $J_z=6$ and not with $J_z=8$. Tab. \ref{table_bos} can be used assuming the system as effective spin $J=6$. However, the suppression feature of Sec. \ref{sec:supp} does not take place anymore. According to our table, GSS had indeed to be expected. 

Another experiment\cite{Donati2016} is Ho on MnO. Here, $\chi = 4$ while spin magnitude is found to be $J = 8$. Also in this case \textit{ab-initio} calculations reveal the presence of a groundstate level crossing. The ratio between the uniaxial anisotropy term and the biggest transversal anisotropy term is found to be as big as $5\%$. The latter term favours a groundstate occupation of the spin state with $|J_z|=7$, rather than $|J_z|=8$. With the prescriptions above indicated, Tab. \ref{table_bos} can still be exploited (using $J=7$) and protection from GSS and SE switching are found, in agreement with the statements of the authors.

A similar situation happens in a third experiment. Dy atoms are deposited on graphene\cite{Baltic2016}. Hence $\chi = 6$ and $J=8$. Again, a strong uniaxial field leads to a groundstate occupation of the spin state $|J_z|=7$. The authors found protection from GSS and SE switching, which agrees to the indication of Tab. \ref{table_bos} (using $J=7$).

This comparison with real experiments shows that level crossing is likely to happen. When this is case, the groundstate quasi-spin can not be inferred from the spin magnitude (and $\chi$) only. Nonetheless, as shown above, our theory can still be applied, for a deep understanding of the system properties, if additional independent informations, e.g. from \textit{ab-initio} calculations or direct measurements, give access to the groundstate quasi-spin.

\section{multiatom cluster systems}
Since not only single-atom nanomagnets but also multiatom clusters are under the attention of researchers\cite{Khajetoorians2013,Loth2012,Gambardella2003,Delgado2016}, we generalize the single atom results to non-frustrated multiatom configurations.

\subsection{Model}

We assume that the atoms interact through Heisenberg-like couplings due to e.g. direct ferromagnetic exchange or indirect Ruderman-Kittel-Kasuya-Yosida interaction\cite{RKKY1954,Zhuo2010}. For simplicity, we do not include Dzyaloshinsky-Moriya interactions\cite{DM1957}. As they might play a role when dealing with rare-earth adatoms and in general with systems with broken inversion-symmetry \cite{Bode2007}, their inclusion is left to future investigations. Thus, the total Hamiltonian 
\begin{flalign} \label{Hmulti}
H_{A} =& \sum_i \left[ H_{A}^{(0)}(i) + H_{A}^{(1)}(i) + \vec B_{i} \cdot \vec J(i)\right] \nonumber \\
& + \sum_{i>j} H_{A}^{int}(i,j)
\end{flalign}
includes the uniaxial anisotropy felt by the $i-$th atom
\begin{align}
H_{A}^{(0)}(i)  = & - \vert D_i \vert \,J^2_{z}(i),
\end{align}
further anisotropy terms $H_{A}^{(1)}(i)$, and the multiatom Heisenberg interaction
\begin{align}
H_{A}^{int}(i,j)= & G_{ij} \,J(i) \cdot J(j).
\end{align}
The effective interaction between the electrons in the metallic surface and the atoms is 
\begin{align} \label{Ht2}
 H_t = \sum_{l} \kappa_{l} \, J(l) \cdot j_{x_l}
\end{align}
where $j_{x_l}=  c^{\dagger}_{x_l} \sigma c_{x_l} \propto \sum_{k,k'} e^{i\,(k-k')\cdot x_l} c^{\dagger}_{k} \sigma c_{k'} $ is the effective spin degree of freedom of the metal electrons coupled to the atom at position $x_l$.

To avoid magnetically frustrated configurations, we restrict the discussion to clusters where one can distinguish two groups of atoms, say $A$ and $B$, such that they have intragroup ferromagnetic coupling ($G_{ij}<0$ if the $i-th$ and the $j-$th atoms are in the same group) and intergroup antiferromagnetic couplings ($G_{ij}>0$ if the $i-th$ and the $j-$th atoms are in different groups). A part from this restriction, the clusters are not required to have other additional properties like, for instance, a specific symmetric spatial configuration of the adatoms that compose it.

\subsection{Operators}

Similarly to $R$ in Eq. \eqref{rotdef}, the rotation generators for every atom may be defined as $R(l)=\exp\{i\, J_z(l)\,2 \pi/\chi \}$. We define the operator associated to the rotation of all spins as
\begin{align} \label{Rtot}
R_{tot} = \otimes_l R(l)= \exp\{i\,J_{z,tot}\,2 \pi/\chi \}
\end{align}
where $J_{z,tot}=\sum_l J_z(l)$ is the projection along the $z$-axis of the total spin.

The mirror operators $M(l)$ at mirror planes by each atom may be defined analogously. 

The time-reversal operator is also trivially generalized to act on multiple spins.

\subsection{Groundstate splitting for $H_t=0$} \label{GSSmulti}

As a first step, we show that a \textit{quasi}-spin can be associated to the groundstates of the multiatom configuration.

With $H_{A}^{(1)}(i)=H^{(int)}_A(i,j)=\vec B_i=0\; (\forall i,j)$, the non-interacting groundstates of the system are products of the groundstates of every independent atom. For instance, with only two atoms, the four groundstates are $| \pm J_{1}\rangle\,| \pm J_{2}\rangle$, $J_i$ being the magnitude of the spin of the $i$-th atom.

We now switch on adiabatically all the interactions $H_{A}^{int}(i,j)$. These terms have actually a higher symmetry than $C_\chi$, namely they are isotropic, and preserves $J_{z,tot}$. Since the non-interacting groundstate has high degeneracy, at first sight it is not clear a priori which states remain groundstate of the system after the switching process.
However, such clusters seam to have the following, \textit{per se} interesting, feature: \\

Conjecture.  \textit{ Given the Hamiltonian in Eq. \eqref{Hmulti} with vanishing $H_{A}^{(1)}(i)$, the groundstate is an eigenstate of $J_{z,tot}$, with eigenvalue in modulus equal to $|J_A -J_B|$, where $J_{A(B)} \vcentcolon =  \sum_{i \in A(B)} J(i)$. By TR symmetry, the groundstate is doubly degenerate if $J_A \neq J_B$.}\\

Through the analysis of the spectrum of several $H_A$ and numerical simulations  (see Sec. \ref{NumMulti}), we got evidence that this conjecture\cite{Nachter2015} holds true. We are able to give a rigorous proof only in first order perturbation theory in the intergroup couplings of the matrix $G$  (the intragroup couplings being allowed to have arbitrary magnitude). This regime is enough to understand how the single-atom features, found in Sec. \ref{sec:satom}, appear also in the multiatom case. Notice that purely ferromagnetic configurations fall into the range of our proof (as either group $A$ or $B$ is empty). Due to the technical character of the proof, we present it in App. \ref{AlmTheo}.

The Marshall theorem, in the generalized fashion by Lieb and Mattis \cite{Lieb1962}, ensures that, at $H^{(0,1)}_A= \vec B_i=0$, for each $l\geq |J_A-J_B|$, the lowest Hamiltonian eigenvalue with total spin magnitude $J_{tot}$ equal to $l$ is a monotone increasing function of $l$ while, for $l\leq |J_A-J_B|$, it is monotone decreasing. Lieb and Mattis have proven that a magnetic field, proportional to $J_{z_i}$, destroys this order. Our conjecture regards the same kind of systems but with an additional finite and negative definite TR symmetric term, the uniaxial anisotropy (also higher order negative definite uniaxial terms may be added). The magnitude of the total spin is not anymore a good quantum number and the ordering of levels is destroyed. Still, according to our conjecture, the groundstates have the property
\begin{align} \label{multiGS}
\vert J_{z,GS} \vert = \vert J_A-J_B \vert
\end{align}
and, crucially, we can associate them well defined \textit{quasi}-spins. The latter are inferred by their eigenvalue under $R_{tot}$ (see Eq. \eqref{Rtot}) and are computed via Eq. \eqref{qmom} inserting $J_z$ according to Eq. \eqref{multiGS}. 

As a further step in the discussion upon the presence of GSS, we switch on the $H_{A}^{(1)}(i)$ terms. As in Sec. \ref{sec:GSSsingle}, if we assume these terms to be small enough such that the initial groundstates are not crossed (in energy) by other levels, then the groundstates \textit{quasi}-spins are preserved. At this point the discussion about the GSS is identical to one done for the single-atom case: when the groundstates \textit{quasi}-spins are integers and are at the TR invariant points of the Brillouin Zone, then GSS takes place. Notice that, according to the conjecture, equal-spin dimers have zero $J_{z,tot}$ (and \textit{quasi}-spin) and their groundstate is generically non-degenerate. We conclude that dimers present GSS even with vanishing $H_{A}^{(1)}(i)$ terms. 

\subsection{Single-electron switching process at $H_t \neq 0$}

We now switch on the small interaction with the metal. Similarly as before (cf. Eq. \eqref{gamma})
\begin{align} \label{gamma2}
\Gamma_{ab} &= \frac{2\pi}{\hbar} \sum_{\mu,\nu} | \langle  \psi_a,\nu | H_t | \psi_b,\mu\rangle|^2 \, e^{-\beta E_\mu} \delta (x) \nonumber \\
& = \frac{2\pi}{\hbar} \sum_{\mu,\nu} \left\vert \langle  \psi_a,\nu |\!\!\!\!\!\!  \sum_{\substack{i \\ s\in \{+,-,z\} }} \!\!\!\!\!\!\kappa_{i} J_{s}(i) \cdot j_{x_i\bar s}| \psi_b,\mu\rangle \right\vert^2\! e^{-\beta E_\mu} \delta (x) \nonumber \\
&= \frac{2\pi}{\hbar} \sum_{\mu,\nu} \left\vert \vec\kappa \cdot\vec V \right\vert^2 \, e^{-\beta E_\mu} \delta (x) 
\end{align}
where $\left(\vec V\right)_i = \sum_{s\in \{+,-,z\} }  \langle  \psi_a| J^{(i)}_{s} | \psi_b \rangle \, \langle  \nu| j_{x_i\bar s}| \mu \rangle $, $\left(\vec \kappa\right)_{i} = \kappa_i$ and $x=E_{\nu}-E_{\mu}+E_{a}-E_{b}$.\\
$\Gamma=0$ only when $\vec V \cdot \vec {\kappa} = 0$ for all possible $\mu,\nu$ states i.e. when $\langle  \psi_a |\,J^{(i)}_{s}| \psi_b \rangle$ are vanishing for every $i$. Fortunately, an analog of Eqs. \eqref{selrule} and \eqref{TRSconstraint2}, with $J_s$ replaced by $J^{(i)}_s$, does hold and, in particular we get again protection from SE switching process for integer spin system.
The protection here may be subtle. Consider, for instance, a system with $\chi=6$ made up of two atoms with spins $J=7/2$. If their coupling $G$ is ferromagnetic, the total spin is $J=7$ and the system presents no SE switching process, according to Eq. \eqref{gamma2} and Tab. \ref{table_bos}. In particular, this fact holds true even when the atoms are set at big reciprocal distance. However, in this situation the two atoms may be regarded as non-interacting and present individually SE switching processes, according to Tab. \ref{table_fer}. We remark that there is no contradiction between the two viewpoints: the full groundstate, being a product of the groundstates of the two atoms in the non-interacting limit, needs two electrons to be fully switched. Even though quantitatively, the dimer has a big rate of switching, qualitatively it remains SE switching protected. 

We warn the reader that switching transitions between degenerate groundstates of integer spin systems can be observed. However, these transitions must be attributed to $2n$-electrons processes, with $n$ integer, (as one can see generalizing Eq. \ref{TRSconstraint2}) and not to single-electron ones\cite{note9}.

Finally, we notice that the suppression feature of Sec. \ref{sec:supp} is not present for the multiatom case. The difference with the single-atom case lies in the fact that the state $|1+J\rangle$ was a forbidden state there, while here its analog, $\vert1+|J_A-J_B|\rangle$ is, in general, allowed.

\subsection{Numerical simulations} \label{NumMulti}
\begin{figure}[tb]
\centerline{\includegraphics[width=1\linewidth]{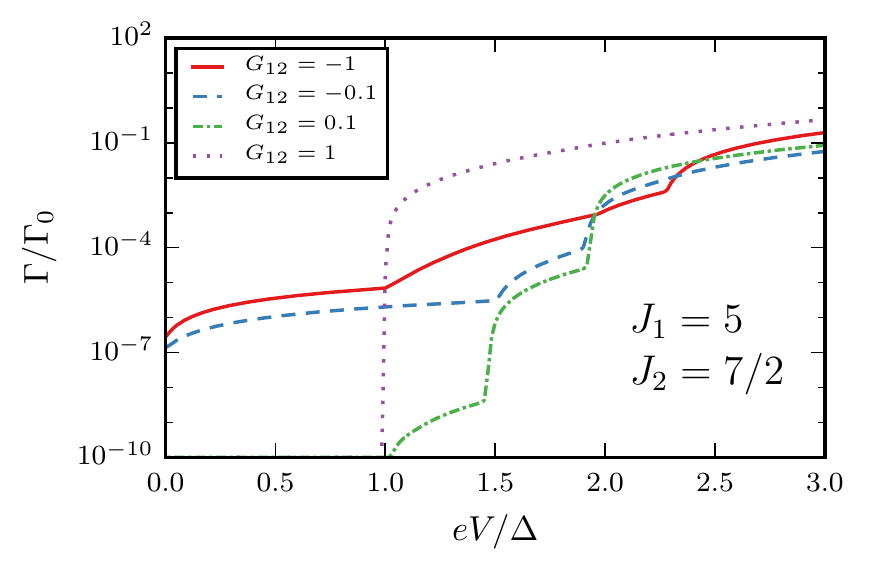}}
\caption{Bias-dependent switching rate of a dimer with spin magnitudes $J_1$ and $J_2$ in a six-fold symmetric crystal ($\chi=6$), at different strength of the exchange coupling $G_{12}$ (in unit of $D$). The tip is placed on top of the first atom i.e. $\kappa^2_1/D=0.1$ and $\kappa^2_2/D=0$. Here, $\alpha_6^6/D= 1\cdot10^{-3}$, all other parameters are as in Fig.~\ref{fig:bias}}.
\label{fig:Multi1}
\end{figure}

\begin{figure}[tb]
\centerline{\includegraphics[width=1\linewidth]{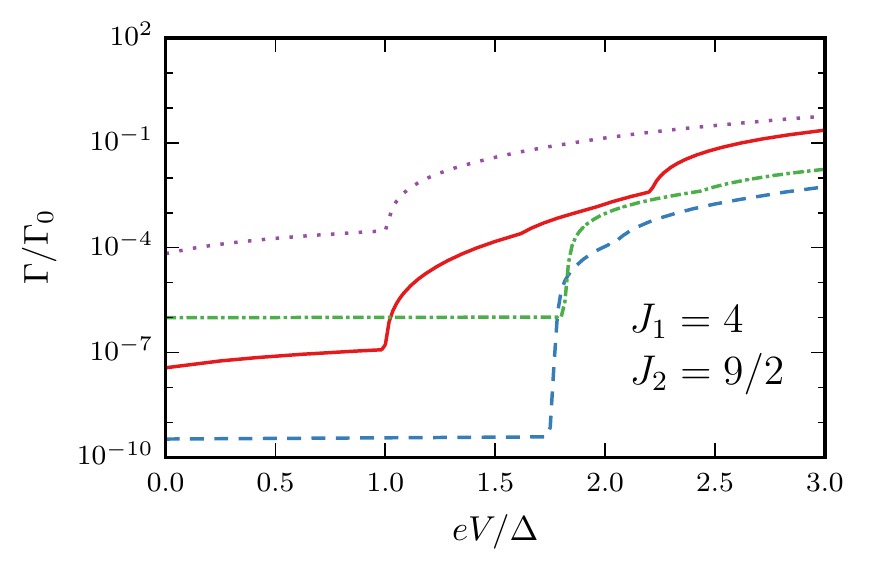}}
\caption{Same as in Fig.~\ref{fig:Multi1} but for a dimer with different spin magnitudes.}
\label{fig:Multi2}
\end{figure}

We perform numerical simulations similar to the ones shown in section \ref{num1}, focusing only on the bias dependency of the switching rate. We analyze the cases of two dimers with same \textit{quasi}-spins when they are in a ferromagnetic configuration but different when in a antiferromagnetic one (see Figs.~\ref{fig:Multi1},\ref{fig:Multi2}). Since we are interested only in the stability features, we assume vanishing distance between the atoms. 

When the coupling is ferromagnetic ($G_{12}<0$), both dimers are predicted to be unstable, as in both cases $\vert J^{(q)}_{GS}\vert = 5/2$. Both our simulations confirm the expectation. The case $G_{12}=-0.1$ in Fig.~\ref{fig:Multi2} points to an important feature of multiatom configurations: the rate (at zero voltage) can be very small. Notice that, in order to get rates $\Gamma$ comparable with the single-atom case, we need to increase the transversal anisotropy ($\alpha_6^6/D$) about two orders of magnitude. 

 When the coupling is antiferromagnetic ($G_{12}>0$) the case in Fig.~\ref{fig:Multi1} is predicted to be stable, as  $\vert J^{(q)}_{GS}\vert = 3/2$, while the other one unstable, as $\vert J^{(q)}_{GS}\vert = 1/2$.

Notice that the cases $G_{12}=\pm0.1$ in Fig.~\ref{fig:Multi2} present the first kink at higher voltage than the one which corresponds to the first excitation energy ($\Delta$). This interesting phenomenon is a prerogative of multiatom systems (with $\chi=6$): the first excited states can be not SE-connected to the groundstates. When it happens, the transition rates between these states are suppressed and a new channel of switching opens only at higher voltage when second excited states can get excited. This feature may be exploited to increase the energy-window of stability (in units of $\Delta$). For instance, a dimer with $J_1=4$ and $J_2=2$ with the same parameter set as in the figures and antiferromagnetic coupling $G_{12}=-0.1$ has groundstate \textit{quasi}-spin  $\vert J^{(q)}_{GS}\vert = 2$ while the first excited states have $\vert J^{(q)}_{GS}\vert = 0$. The groundstates are then SE-switching protected and the switching (at small $T$) activates only at $eV \sim 2\Delta$ in correspondence with the second excited states.

As a final remark, we see that our numerical simulations support the conjecture in Sec. \ref{GSSmulti}. Indeed, the cases with $G_{12}=1$ fall outside the range of validity of our proof (see App. \ref{AlmTheo}), but the numerics confirms our expectations in terms of the stability of the groundstates.

\subsection{Discussion}

Clusters seem to behave as single atoms as far as our analysis is concerned. We can associate them a \textit{quasi}-spin and they have analogous selection rules for SE switching processes. A difference with the single-atom case is that the magnitude of total spin of the groundstates is not well defined anymore (a part in the ferromagnetic case). Nonetheless, this is of no consequence since the unique quantum number needed to determine the symmetry selection rules is the \textit{quasi}-spin. 

One other caveat is that the feature of missing SE switching process for small spin systems (see Sec. \ref{SESsingle}) is not present here unless for all atoms, that compose the cluster, $\chi>2J>1$ holds. These systems do not follow our tables but could be addressed separately as they are relatively simple to be studied. Moreover, also the suppression feature of Sec. \ref{sec:supp} is not present.

To conclude, we inform that Tabs. \ref{table_bos} and \ref{table_fer} can be used for the multiatom case. However, the spin magnitude of the single atom has to be replaced with an effective groundstate spin magnitude $|J_A-J_B| + \chi$, where the ``$+\chi$" term is conveniently added to avoid those small spins constraints, as illustrated above.

\section{Summary and Outlook} 

We focused on the dynamic properties of generic nanomagnets made of absorbed adatoms on metallic or insulating surfaces. We presented a complete and comprehensive discussion on the implications of the symmetries of the system on the stability of the magnetic states. In particular, the symmetries of interest are the rotational, the mirror and the time-reversal symmetry. All our results are summarized in Tabs. \ref{table_bos} and \ref{table_fer}. Given the effective spin magnitude of the adatoms and the symmetries of the system, our main results, the tables (Tab. \ref{table_bos}, \ref{table_fer}), indicate whether a nanomagnet is stable by its desirable properties: absence of groundstate splitting and single-electron switching processes.  
Further, we discovered the interesting feature of suppression of single-electron switching process in some systems with uniform and weak transversal anisotropy.

Finally, we presented an extension of our symmetry considerations to a rather generic class of multiatom clusters. The tables (Tabs. \ref{table_bos}, \ref{table_fer}) can still be used if the effective spin magnitude of each adatoms composing the cluster is known. Here, we limited our study to generic non-frustrated configurations. Our analysis of the multiatom clusters could be in future extended to many other symmetries (for example to systems where the adatoms form chains or lattices).

All our results are supported by numerical simulations which show the switching behavior of these nanomagnets and offer guidance for experimental measurements, e.g. by scanning tunneling microscopy.

We notice that high rotational symmetry is desirable for the stability of nanomagnets. Indeed, the Brillouin zone associated to the adatom or cluster eigenstates has many elements and systems with a big difference between the grounstates' \textit{quasi}-spin can be found.

We found that the mirror symmetry does not influence qualitative results.

As one rules out the translation symmetry of the substrate, $\chi$ is not restricted anymore by the crystallographic restriction theorem\cite{Bamberg2003}. However, our expressions, been generic, are still valid and applicable. For instance, if a single adatom is put at the high symmetric point of a pentagonal \textit{quasi}-crystal, our expressions apply with $\chi=5$ and we expect the system to have similar (but richer) properties compared to a system with $\chi=3$. Moreover, the adatom could be put on top of an high symmetric molecule with $\chi>6$\cite{Wegner}. However, a quantitative analysis that ensures that environmental crystal field (the one due to the support of the molecule) is negligible must be attached to the study.

Future work may be done in this direction or to prove the conjecture in section \ref{GSSmulti} at arbitrary Heisenberg intergroup couplings.

\section{Acknowledgements} 

We have benefited from discussions with C. L. Kane, T. Neupert, D. Wegner and A. A. Khajetoorians.

This research was supported by the Foundation for Fundamental Research on Matter (FOM), the Netherlands Organization for Scientific Research (NWO/OCW), an ERC Synergy Grant and fundings through SFB925, GrK1286.

\appendix
  
\section{Matrix representation of the Hamiltonian in the single-atom case} \label{HamRep}
Here, we analyze an explicit matrix representation of $H_A$ in the single-atom case. This is an alternative to the most straightforward Stevens operator expansion presented in the main text. It proves to be useful for finding the weak constraints on SE switching due to the mirror symmetry and for checking calculations done with other approaches. It may be used for statistical analysis of the system with the tools of Random Matrix Theory\cite{Metha,Marchenko}. 

As in Sec. \ref{sec:stev}, we start considering the spatial symmetry constraints, then we show the one due to TR symmetry.\\

\paragraph{Rotational symmetry.}

The symmetry $[R,H]=0$ imposes all matrix element between different elements with different $r$ to be zero. Clearly the unspecified $H$ can be represent in an hermitian block diagonal form which has, in general, $3$ kinds of blocks: blocks associated to $R$-eigenspaces with real eigenvalue $r$ and pairs of blocks associated to eigenspaces with conjugated pairs of eigenvalues $r$. To simplify the discussion, assume one real $r$ block, call it $\mathcal{Q}$, and one pair of blocks, call them $\mathcal{X}$ and $\mathcal{Y}$, then:
\begin{align} \label{Hstruc1}
H &= \begin{bmatrix} 
     \mathcal{Q} & \multicolumn{1}{|c}{0} & 0 \\ \cline{1-2}
     \multicolumn{1}{c|}{0} & \mathcal{X} & \multicolumn{1}{|c}{0} \\ \cline{2-3}
     0& \multicolumn{1}{c|}{0} & \mathcal{Y}  
    \end{bmatrix}
\end{align}

\paragraph{Mirror symmetry.}

When acting on the spin eigenbasis $\{|j_z\rangle\}$, the mirror operator in Eq. \eqref{Rm} can be written as
\begin{align} \label{Rmonspin}
M =
\begin{cases}
A & \mathrm{for\;integer\; spin}\\
iA & \mathrm{for\;half\;integer\; spin}
\end{cases}
\end{align}
with $A$ a matrix with antidiagonal filled with ones and zeros outside.

The Hamiltonian elements get the simple constraint:
\begin{align} \label{Rrep}
\langle j_z |H|j_z\rangle &= \langle -j_z|H|-j'_z\rangle. 
\end{align}
It is convenient, to order the elements of this basis in each block by putting states with descending order in $j_z$, for blocks $\mathcal{Q}$ and $\mathcal{X}$, and with ascending order for $\mathcal{Y}$. For instance, with $J=3$ and $\chi=3$ such basis is $\{ |j_z\rangle\} = \{|3\rangle,|0\rangle,|-3\rangle,|2\rangle,|-1\rangle,|-2\rangle,|1\rangle\}$. This choice will be particularly useful when we will implement the TR symmetry.

We see clearly that the mirror symmetry creates a constraint between the elements of block $\mathcal{Q}$ and implies that the block $\mathcal{X}$ must be equal to the block $\mathcal{Y}$. \\

\paragraph{Time reversal symmetry.}

We show the constraint due to TR symmetry alone; spatial symmetries are not necessarily present. We order the states of the spin eigenbasis such that TR-partners are grouped together. For instance, with $J=3$ and $\chi=3$ such basis is $\{ |j_z\rangle\} = \{|3\rangle,|-3\rangle,|2\rangle,|-2\rangle,|1\rangle,|-1\rangle,|0\rangle\}$. In this basis the operator $T$ is represented as
\begin{align} \label{Trep}
T=K\;\oplus_{j\neq0}^J \left[\sigma_x^{(j)} \cos(\pi\,j)+i\sigma_y^{(j)} \sin(\pi\,j)\right] \nonumber\\ \oplus
\begin{cases}
 1^{(j=0)}\, & \mathrm{for\;integer\; spin}\\ 
-- &  \mathrm{for\; half\;integer\;spin}
\end{cases}
\end{align}
where the superscript $(j)$ indicates that the operator acts on the time reversal pair $\{|j \rangle,|-j \rangle\}$ (or on the singlet state when $j=0$).
For the sake of the discussion, we discard the presence of the $J_z=0$ state for integer spin systems; we reintroduce it next paragraph. The TR symmetry constraint reads 
\begin{align} \label{TRconstr2}
\bar h_{lk} = 
\begin{cases}
 (-1)^{l+k} \sigma_x \,\bar h^*_{lk}\, \sigma_x  & \mathrm{for\;integer\; spin}\\ 
(-1)^{l+k-1} \sigma_y  \,\bar h^*_{lk}\, \sigma_y  &  \mathrm{for\; half\;integer\;spin}
\end{cases}
\end{align}
here all $\bar h_{lk}$s are $2 \times 2$ Hamiltonian submatrices acting on time reversal pairs with $|j_z|=l,k$. \\
We see that, for integer systems,
\begin{align} \label{Itrs}
\bar h_{lk} = 
\begin{cases}
\begin{pmatrix}
 a & b \\
 b^* & a^*
\end{pmatrix}, \qquad  \mathrm{for\;l+k\;even}\\ 
\begin{pmatrix}
 a & b \\
 -b^* & -a^*
\end{pmatrix}, \qquad  \mathrm{for\;l+k\;odd}\\ 
\end{cases}
\end{align}
For half-integer systems
\begin{align} \label{HItrs}
\bar h_{lk} = 
\begin{cases}
\begin{pmatrix}
 a & b \\
 b^* & -a^*
\end{pmatrix}, \qquad  \mathrm{for\;l+k\;even}\\ 
\begin{pmatrix}
 a & b \\
 -b^* & a^*
\end{pmatrix}, \qquad  \mathrm{for\;l+k\;odd}\\ 
\end{cases}
\end{align}

\paragraph{General form with all symmetries.}
When TR symmetry is added to the spatial symmetries, the Hamiltonian structure in Eq. \eqref{Hstruc1} becomes
\begin{align} \label{Hstructure}
H   &=\begin{cases}
     \begin{bmatrix}
       P & \multicolumn{1}{|c}{0} & 0 \\ \cline{1-2}
     \multicolumn{1}{c|}{0} & S & \multicolumn{1}{|c}{0} \\ \cline{2-3}
     0& \multicolumn{1}{c|}{0} & S  
    \end{bmatrix}, &\text{for $\chi\neq 3$}\\
    \\
     \begin{bmatrix}
      C^{\dagger} P C& \multicolumn{1}{|c}{0} & 0 \\ \cline{1-2}
     \multicolumn{1}{c|}{0} & C^{\dagger}S C & \multicolumn{1}{|c}{0} \\ \cline{2-3}
     0& \multicolumn{1}{c|}{0} & C^{\dagger}S C 
    \end{bmatrix},&\text{for  $\chi=3$}\\ 
    \\
    \end{cases}  
\end{align}
where $P$ is a real matrix where the superdiagonals have components disposed in a palindromic way\cite{note7}; $S$ is a symmetric matrix; $C=\mathrm{diag}\{1,i,1,i,\dots \}$ where the alternating pattern is limited by the dimension of the block. Notice that block $\mathcal{Q}$ is not present for half-integer spin systems with $\chi\neq3$ (hence $P$ is null), since there are not TR invariant \textit{quasi}-spins in the BZ. 

We remark that, for $\chi\neq 3$, the eigenvectors can be chosen to be real, since the Hamiltonian matrix is real and symmetric. For $\chi = 3$, the eigenvectors are complex but can be written in the form $\vec w = C^{\dagger} \vec v$ with $\vec v$ a real vector. In Dirac notation, the eigenstates could be written as
\begin{align} \label{Hvector}
|\psi \rangle =\begin{cases}
\sum_{j\in {block}} v_j  |j\rangle, & \quad \text{for $\chi\neq 3$} \\
\sum_{j\in {block}} c_{jj}v_j  |j\rangle  & \quad \text{for $\chi= 3 $}
\end{cases}
\end{align}
Hermiticity constraints the diagonal elements of the half-integer cases bringing to Kramers degeneracy. One relevant consequences of this fact is that systems with $J=3/2$ and $\chi=3$ are protected from SE switching processes (as indicated in Tab. \ref{table_fer}).

\section{Appendix B. Weak constraints on the SE switching processes} \label{Rmconstraint}

Here, we show the constraints to the quantity
\begin{align} \label{AppBfocus}
\langle \psi_{GS} | J_{s} |\tilde  \psi_{GS}\rangle,\;(s=+,-,z)
\end{align}
coming from the mirror symmetry and the symmetry under the operator $TM$, effective in a specific regime. The analysis is restricted to the single-atom case. As these constraints appear to affect the SE switching rates only quantitatively we call them ``weak" as opposed to the constraints due to time reversal and rotational symmetries. We do not generalize them to the multiatom case as we expect, also for this case, similar weak constraints.\\

\paragraph{Constraint from the mirror symmetry.}
Consider the quantity in the expression \eqref{AppBfocus} when the mirror symmetry is present. 
The Hamiltonian eigenstates $|\psi\rangle$ can be chosen to be also eigenstates of $R$, since $[H,R]=0$.  The commutation relation $R M = M R^{\dagger}$, then, implies $R \left( M |\psi \rangle \right) = r^* M|\psi \rangle$. This means that $M|\psi \rangle$ is an eigenstate of $R$ but with different \textit{quasi}-spin if $r$ is non real. On the other hand $M|\psi \rangle$ and $|\psi \rangle$ must have the same energy since $[H,M]=0$. Therefore, when $r$ is not real $M|\psi \rangle \perp |\psi \rangle$ i.e. $M|\psi \rangle = a |\tilde \psi\rangle \vcentcolon =  a T|\psi \rangle$, with $a$ a unit complex number. Applying $M$ to both sides of the previous equation and using $M^2=\pm 1 $, after a trivial manipulation one gets $M|\tilde\psi\rangle =  \pm a^*|\psi \rangle$, where plus(minus) sign refers to integer(half integer) spin systems. About $a$ we only need to know whether it is real or imaginary, as it will be clear in a moment. From Eq. \eqref{Trep} and the specification of the form of $|\psi\rangle$ in Eq. \eqref{Hvector}, we see that $T$ maps the vector $v$, for $\chi\neq 3$, in another real vector, and $w=C^\dagger v$, for $\chi=3$, to the vector $C^\dagger v'$ (with $v'\neq v$). Differently, $M$ maps the vectors to same-shape vectors but multiplied by the imaginary unit for half-integer spins (see Eq. \eqref{Rmonspin}). Therefore, $a$ is real(imaginary) for integer(half-integer) spin systems. We are now ready to obtain the SE switching constraint:
\begin{align} \label{Rmpassages}
\langle \psi_{GS} | J_{\pm} |\tilde \psi_{GS}\rangle = &\langle \psi_{GS} | M^{\dagger} J_{\mp} M |\tilde \psi_{GS}\rangle \nonumber\\
= & \pm (a^*)^2 \langle \tilde \psi_{GS} | J_{\mp} |\psi_{GS} \rangle \nonumber\\
= &\pm (a^*)^2 \langle \psi_{GS} | J_{\pm} |\tilde \psi_{GS} \rangle^*\nonumber\\
= & \langle \psi_{GS} | J_{\pm} |\tilde \psi_{GS}\rangle^*
\end{align}
where the external plus(minus) sign refers to integer(half integer) spin systems.

Finally, we conclude
\begin{align} \label{Rmconstr1}
\mathrm{Im} \langle \psi_{GS} | J_{\pm} |\tilde \psi_{GS}\rangle = 0.
\end{align}
When $r$ is real, it is of interest to consider whether there is a constraint on $\langle \psi_{GS} | J_{z} |\tilde \psi_{GS}\rangle$, for half-integer spin systems (then with $\chi=3$). We show first that 
\begin{align}
\langle \psi | M| \psi\rangle = 0.
\end{align}
Using Eq. \eqref{Hvector}, we can rewrite the the l.h.s of the previous equation as the scalar product $\left(w, M w\right)=\left(C^\dagger v, M C^\dagger v\right)$. Remember, now, that $M=iA$ and notice that the dimension of the block $\mathcal{Q}$ must be even, therefore $iA C^{\dagger}=C A$ holds. The quantity, then, simplifies to $\left( v, C^2A v\right)$ which vanishes since $v$ is real and $C^2 A$ antisymmetric. Similarly as when $r$ is non-real, we conclude that $M| \psi\rangle = b | \tilde \psi\rangle$.\\ 
One could show that $b$, like $a$ is real(imaginary) for integer(half-integer) spin systems and, with similar passages as before, conclude
\begin{align} \label{Rmconstr2}
\mathrm{Re} \langle \psi_{GS} | J_{z} |\tilde \psi_{GS}\rangle = 0.
\end{align}
Notice that the constraints \eqref{Rmconstr1} and \eqref{Rmconstr2} are not enough to make SE switching processes vanish since, respectively, the real and imaginary parts are left unconstrained and, unfortunately, they are different from zero, given a generic systems.\\

\paragraph{Constraint from the $TM$ symmetry operation.} 
Here, we show the weak constraint on the  expression \eqref{AppBfocus} coming from the symmetry operator $TM$, relevant when the time reversal symmetry is broken by a (rotational symmetry preserving) magnetic field along the $z$ axis. In this situation, the groundstate is non degenerate. However, for small enough $B_z$, the two lower energy eigenstates retain the same quasi spins and eigenvalues under the action of $TM$ as the ones of the two groundstate at $B_z=0$. Calling (improperly) these two lower eigenstates  $| \psi_{GS}\rangle$ and $|\tilde \psi_{GS}\rangle$ one can find:
\begin{align} 
\begin{cases}
\mathrm{Im}\langle \psi_{GS} | J_{z} |\tilde \psi_{GS}\rangle=0\\ 
\mathrm{Re}\langle \psi_{GS} | J_{\pm} |\tilde \psi_{GS}\rangle=0 &\mathrm{for\;integer\; spin} \\
\\
\mathrm{Re}\langle \psi_{GS} | J_{z} |\tilde \psi_{GS}\rangle=0\\ 
\mathrm{Im}\langle \psi_{GS} | J_{\pm} |\tilde \psi_{GS}\rangle=0 &\mathrm{for\;half\;integer\; spin.}
\end{cases}
\end{align}

We limit ourselves to just show this result because its proof is lengthy and the result is just weak constraints which are not enough to make SE switching processes vanish. The reader may appreciate how, at $B_z=0$, these constraints plus the constraints in Eq. \eqref{Rmconstr1} and \eqref{Rmconstr2} imply the time reversal one in Eq. \eqref{TRSconstraint2}.

\section{Appendix C. Prove of the conjecture in Sec. \ref{GSSmulti} at small intragroup couplings} \label{AlmTheo}

We show a proof of the conjecture that appears in Sec. \ref{GSSmulti}, restricted to the case when intragroup couplings of the matrix $G$ are small in comparison to all other energies in $H_A$.

At zeroth order in the intergroup terms in $H^{int}_A$, without uniaxial anisotropy and magnetic field but with finite intragroup terms, the groundstates are $(2J_A+1) \times (2J_B+1)$ product states of the form $J^m_{-,A}|GS_A\rangle \otimes J^n_{+,B}|GS_B\rangle$ with $m(n) =0,\dots,2J_{A(B)}$, $\;J_{\pm,X}=\sum_{i\in X} J_{\pm}(i)$ and $|GS_X\rangle$ is the state with all spin aligned up, for $X=A$, and down, for $X=B$. Clearly, once the uniaxial anisotropy is switched on, $|GS_A\rangle \otimes |GS_B\rangle$, along with the other three states obtained by applying the TR operator to the state in either to $A$, to $B$ or to both, remains the unique groundstate. Indeed, they are eigenstates with maximum eigenvalue of both $\sum_{ij} H_{A}^{int}(i,j)$ and $\sum_i H_{A}^{(0)}(i)$.
Then, we add small intergroup coupling terms in $H_{A}^{int}$, small with respect to the other energies involved. It is straightforward to see that configurations in which the spin of the two groups are oppositely aligned i.e. $|GS_A\rangle \otimes |GS_B\rangle$ along with its TR partner, gain a negative first-order perturbation energy. This energy is equal to $-\sum_{i\in A,j \in B} G_{ij} J(i)J(j)$. On the contrary, the other two states (aligned) gain the same term but with opposite sign. Since the intergroup coupling preserves the value of $J_{z,tot}$ of the perturbed states, the new groundstates will have the same $J_{z,tot}$ of $|GS_A\rangle \otimes |GS_B\rangle$ and its TR partner, given by $\pm (J_A-J_B)$. Thus, the conjecture is proven for small intergroup couplings as claimed in the main text.


\begin{thebibliography}{99}
\bibitem{note1} In our convention, $\mathbb{N}$ includes the zero unless further specifications are present.
\bibitem{note4} To see this, let $| \psi \rangle$ be an eigenstate of $M$. Its eigenvalue is a unit complex number (see App. \ref{Rmconstraint} for details). Considering that $\{ J_z,M\}=0$, the magnetization of the state along $z$ satisfies $\langle \psi | J_{z} | \psi\rangle=-\langle \psi | M^\dagger J_{z} M| \psi\rangle=- \langle \psi | J_{z} | \psi\rangle$. Thus, the magnetization is vanishing.
\bibitem{note5} To be precise, for the approximation to be valid, the second-order terms must be smaller then the first order ones. This happen if $\varepsilon$ verifies $\varepsilon \ll 1/(2J \alpha_{\rm max})$, with $\alpha_{\rm max}= \max_k \alpha_k$.
\bibitem{note7} We define superdiagonal of order $m, (m \in \mathbb{Z})$, the vector with component $v_i=H_{ij}$ with $i-j=m$. For palindromic vector we mean, for instance $v=(4,2,-1,9,-1,2,4)$.
\end{thebibliography}
\end{document}